# Element-Specific Solute Trapping and Grain Structure Evolution during Laser Powder Bed Fusion of Multicomponent Alloys

Xinxin Yao [a], James Hanagan[b], Md Shafiqur Rahman Jame[c], Mallikharjun Marrey[d], Mohsen Taheri Andani[c], Raymundo Arróyave[b], Veera Sundararaghavan [e], Lei Chen [a*]

[a] *Department of Mechanical Engineering, University of Michigan, Dearborn, MI 48128, USA*
[b] *Department of Materials Science and Engineering, Texas A&M University, College Station 77843, TX, USA*
[c] *Department of Mechanical Engineering, Texas A&M University, College Station, USA*
[d] *AlphaSTAR Technology Solutions, Irvine, CA 92614*
[e] *Department of Aerospace Engineering, University of Michigan, Ann Arbor, 48109, MI, USA*

**Abstract**: Under the rapid solidification conditions of laser powder bed fusion (LPBF), solute trapping manifests in an element-specific manner, altering nonequilibrium partitioning, constitutional undercooling, and grain selection behavior in multicomponent alloys. Here, we elucidate the mechanisms by which element-specific solute trapping governs nucleation behavior and grain structure evolution during LPBF demonstrated on a SS316L. This requires quantitative description of nonequilibrium multicomponent thermodynamics and grain evolution across broad LPBF solidification conditions, which is achieved through a CALPHAD-informed Gaussian Process Regression (GPR)-assisted Phase-Field (PF) approach. The predicted transitions in grain morphology and grain size are validated against EBSD measurements under multiple LPBF processing conditions. Results demonstrate that increasing solidification rate drives a composition-dependent transition from solute diffusion-controlled nucleation to solute trapping-controlled grain growth, where nonequilibrium solute redistribution intensified by solute trapping suppresses equiaxed grain formation despite high cooling rates. Quantitative decomposition of multicomponent undercooling further reveals distinct element-specific sensitivities to solute trapping, where C, Cr, and Mo remain dominant contributors to the overall undercooling, while the undercooling contribution of low-partitioning elements such as S and P are strongly suppressed relative to their equilibrium values under rapid solidification conditions. These results reveal how element-specific solute trapping governs grain selection in multicomponent alloys, providing a mechanistic basis for alloy design under nonequilibrium solidification conditions.

[*] Corresponding author: Lei Chen, Email: leichn@umich.edu

## 1. Introduction

In laser powder bed fusion (LPBF) additive manufacturing of multicomponent alloys such as 316 L stainless steel, tailoring mechanical properties through precise control of process parameters has been a long-standing objective for the industrial community [1-3]. The mechanical behavior of LPBF-fabricated components is inherently governed by their microstructural characteristics, including grain size, morphology, and anisotropy [4, 5]. Among these features, the nucleation behavior within the melt pool plays a pivotal role in determining the subsequent grain morphologies, such as equiaxed and columnar structures [6]. However, the nucleation behavior remains challenging due to the complex coupling between high thermal gradients, rapid solidification rates, and local solute redistribution, which collectively affect the columnar-to-equiaxed transition (CET) during rapid solidification [7, 8].

Nucleation behavior during solidification is closely related to the local undercooling ahead of the solid–liquid interface [9]. To represent this relationship, conventional Gaussian-type nucleation models is widely employed [10-15], where the nucleation probability is prescribed through a Gaussian distribution of undercooling. Such models rely on empirical fitting parameters, including the mean nucleation undercooling and its standard deviation [11]. While these parameters can be experimentally characterized under conventional casting or slow solidification conditions, their determination becomes challenging under LPBF conditions due to the inherent process complexity and rapid thermal fluctuations [16]. Moreover, physically informed nucleation theories indicate that nucleation behavior does not necessarily exhibit a Gaussian distribution on undercooling [8], suggesting that conventional Gaussian-type nucleation models may not adequately describe nucleation behavior during rapid solidification in LPBF.

In addition, the applicability of conventional nucleation models becomes further limited in multicomponent alloys, where nucleation behavior is strongly influenced by composition-dependent liquidus properties and nonequilibrium partitioning behavior [17]. Conventional analytical approaches rely on simplified descriptions of solute redistribution. For example, multicomponent alloys such as IN718 [13], Ti-6Al-4V [15, 18], and SS316L [19] are approximated as effective binary systems to facilitate analytical treatment, limiting the ability to capture composition-dependent solidification behavior. As a result, key thermodynamic quantities governing solidification, such as partition coefficient, liquidus slope, and constitutional undercooling, are typically evaluated under assumptions that are not valid for multicomponent systems during rapid solidification process [2, 20, 21]. In reality, diffusion behavior is inherently composition-dependent and strongly influenced by interactions among alloying elements, as demonstrated by Andersson et al. [22] and Kattner et al. [23]. Under rapid solidification conditions, limited solute diffusion further promotes element-specific solute trapping, leading to pronounced deviations from equilibrium partitioning behavior. Durga et al. [24] further demonstrated that the effective partition coefficient of each alloying element is highly sensitive to diffusivity under nonequilibrium solidification conditions. These coupled diffusion and solute trapping effects fundamentally alter local solute redistribution and constitutional undercooling behavior during rapid solidification.

Beyond conventional empirical nucleation models, significant effects have been devoted to developing thermodynamically grounded theories for CET prediction in multicomponent alloys. Classical solidification theory provides a physical foundation for describing grain selection behavior during solidification. The pioneering work by Hunt [24] established an analytical criterion linking the onset of equiaxed growth to the undercooling ahead of the solidification front, thereby offering a mechanistic description of CET. Subsequent developments have extended this framework to incorporate more realistic descriptions of dendritic growth. In particular, Gäumann et al. proposed a numerical CET model that integrates dendrite tip undercooling calculated from Kurz–Giovanola–Trivedi (KGT) theory [25], enabling a more accurate representation of the interplay between thermal conditions and dendrite growth kinetics. Building upon this approach, Kurz et al.[26] used this model to calculate a solidification microstructure selection map, which delineates the regimes of planar, columnar, and equiaxed growth as functions of $G$ and $R$. While these models provide a thermodynamically grounded framework for microstructure selection, their extension to multicomponent alloys remains fundamentally limited by the lack of accurate thermodynamic input under nonequilibrium solidification conditions. In this context, CALPHAD provides a rigorous and self-consistent thermodynamic dataset capable of capturing multicomponent effects and supplying the necessary input parameters for CET modeling [23]. This further motivates the development of thermodynamically consistent coupling between CALPHAD-based CET predictions and nucleation behavior in microstructure evolution simulations under rapid solidification conditions.

In the present work, CALPHAD-based CET predictions are incorporated into phase field (PF) simulations to investigate nucleation behavior and grain structure evolution during LPBF solidification of SS316L, as illustrated in Fig. 1. Multicomponent thermodynamic calculations are first performed to describe the evolution of equiaxed grain fraction $\phi$ over a broad range of thermal gradients and solidification rate conditions. The resulting CET behavior is subsequently linked to local nucleation probability $p$ through a physically informed relationship and incorporated into PF simulations to describe grain nucleation, growth, and competitive evolution. The model captures the coupled effects of multicomponent diffusion, nonequilibrium partitioning, constitutional undercooling, and element-specific solute trapping during rapid solidification. To establish continuous mapping between solidification conditions and CET behavior, Gaussian Process Regression (GPR) is employed to interpolate CALPHAD-predicted CET behavior across broad solidification conditions. The predicted grain morphology and grain size evolution are quantitatively validated against EBSD measurements under multiple LPBF processing conditions. We demonstrate the element-specific sensitivity of solute trapping and its critical role in governing nonequilibrium partitioning, constitutional undercooling, and grain selection behavior during rapid solidification.

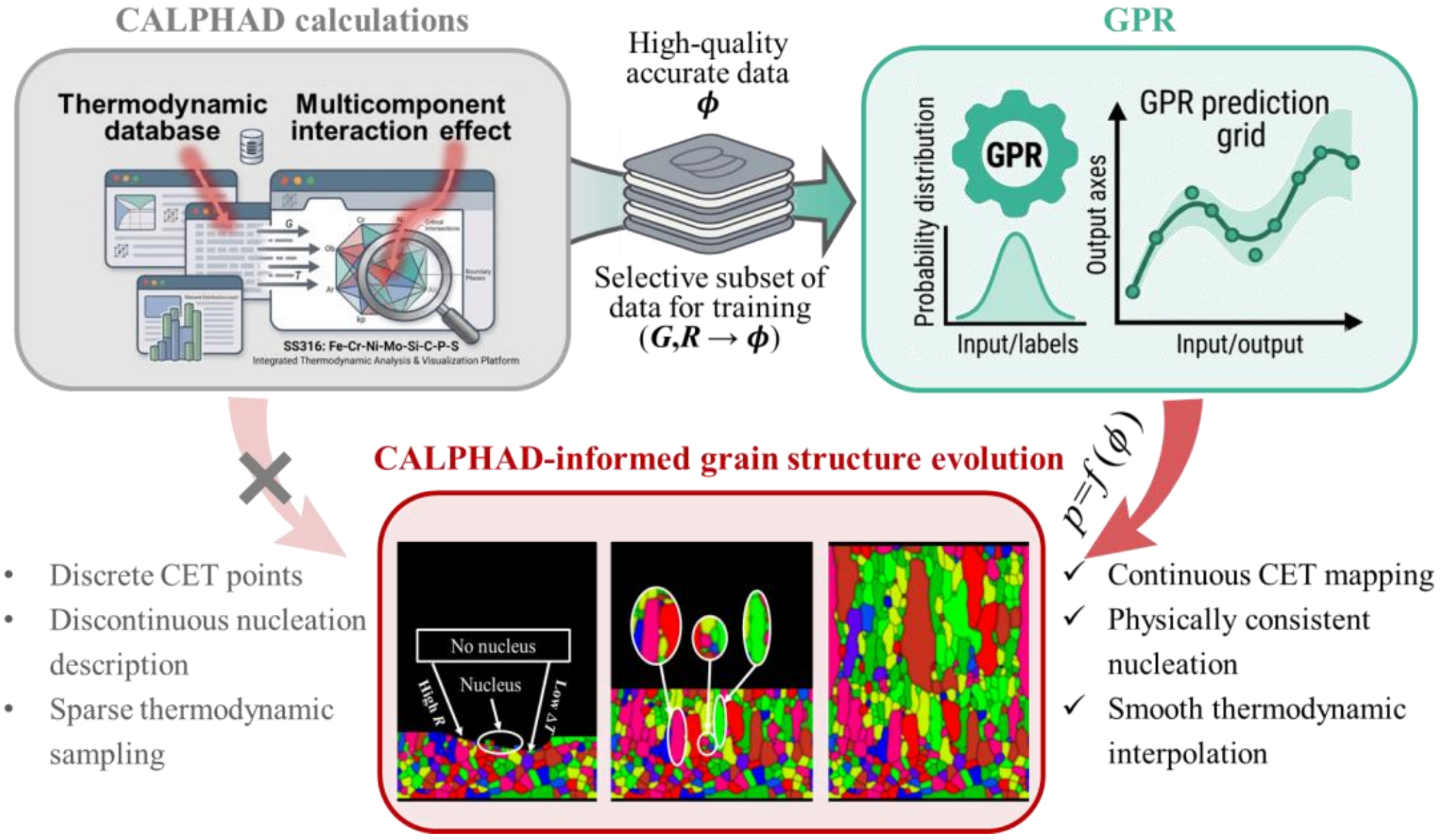

Fig. 1. Schematic illustration of thermodynamic linkage between CALPHAD-based CET predictions and grain structure evolution during LPBF solidification of SS316L.

## 2. Methods

### 2.1 Grain growth via phase field model

PFM has been proven capable of accurately predicting grain nucleation and growth, epitaxial growth, and competitive growth in additive manufacturing process [13, 15, 27-30]. In our previous works, the developed PFM was employed to simulate the single-layer grain structure evolution in IN718 [13] and the multi-layer grain growth behavior in Ti-6Al-4V [15], achieving good agreement with experimental observations. With this theoretical formulation established, the present PFM is applied to predict grain evolution in LPBF SS316L. The governing equation for the grain structure evolution is expressed as:

$$\frac{\partial \eta_q(\mathbf{r},t)}{\partial t} = -L_q(T)\frac{\delta F(t)}{\delta \eta_q(\mathbf{r},t)} \quad (q=1,2,...,Q) \tag{1}$$

where $\{\eta_q\}$ is a set of continuous order parameters between 0 and 1, representing the grain structure at a given time $t$ at each simulation cell $\mathbf{r}$. $L_q$ is grain boundary mobility coefficient related to both temperature and physical property of material. $F$ is the total free energy of this system. Related equations are given as:

$$L_q(T) = L_0(\frac{T}{T_a})^m \exp(-\frac{\Delta Q}{R_g T}) \tag{2}$$

$$F(t) = \int \left[ w f_0(\eta_1(\mathbf{r},t), \eta_2(\mathbf{r},t), ..., \eta_Q(\mathbf{r},t)) + \sum_{q=1}^{Q} \frac{k_q}{2}(\nabla \eta_q(\mathbf{r},t))^2 \right] d\mathbf{r} \tag{3}$$

$$f_0(\{\eta_q(\mathbf{r},t)\}) = -\frac{1}{2}\sum_{q=1}^{Q}\eta_q^2(\mathbf{r},t) + \frac{1}{4}(\sum_{q=1}^{Q}\eta_q^2(\mathbf{r},t))^2 + \frac{1}{2})\sum_{q=1}^{Q}\sum_{s>q}^{Q}\eta_q^2(\mathbf{r},t)\eta_s^2(\mathbf{r},t) \quad (4)$$

where $L_0$ is the pre-exponential coefficient, $T_\mathrm{a}$ is the ambient temperature, $\Delta Q$ is the activation energy of SS316L, and $R_\mathrm{g}$ is the gas constant. $w$ is the barrier height that governs the interface stability, $f_0(\{\eta_q(\mathbf{r}, t)\})$ denotes the local free density associated with each grain structure. $k_\mathrm{q}$ is the gradient coefficient corresponding to the grain boundary of each grain $q$, which is related to the grain orientation and the local heat-flow direction.

2.2 Multicomponent thermodynamic interaction

Equiaxed fraction predictions were queried from Thermo-Calc's CET property calculator with the TC-Python API, utilizing data from Thermo-Calc's TCFE13 thermodynamic database and MOBFE8 kinetic database, both developed for steel alloy. This calculator implements the CET model first proposed by Gäumann et al. and was queried in a logarithmically spaced grid of $G$ (K/m) and $R$ (m/s) solidification conditions. At each pair of $G$ and $R$, the fraction of equiaxed grains was returned and recorded if the solidification conditions were predicted to result in dendritic solidification. If the given conditions were predicted to result in planar solidification, a value of -1 was instead returned to differentiate conditions that do not yield any kind of dendritic solidification. The relationship between $G$, equiaxed grain fraction $\phi$ and total undercooling $\Delta T$ can be expressed as [31],

$$G = \frac{1}{n+1}\sqrt[3]{\frac{-4\pi N_0}{3\ln(1-\phi)}}\Delta T(1-\frac{\Delta T_e^{n+1}}{\Delta T^{n+1}}) \quad (5)$$

where $N_0$ is the density of potential nucleation sites, $n$ is the material-dependent exponent. The critical undercooling $\Delta T_e$ for heterogeneous nucleation dependent on the size of the nucleating site $d$ and is given by,

$$\Delta T_e = \frac{4\gamma}{\Delta S_f d} \quad (6)$$

where $\gamma$ is the interfacial energy, and $\Delta S_f$ is the fusion entropy of per volume.

The total undercooling $\Delta T$ is the sum of constitutional undercooling $\Delta T_c$, thermal undercooling $\Delta T_t$, radial undercooling $\Delta T_r$, and kinetic undercooling $\Delta T_k$, which is expressed as,

$$\Delta T = \Delta T_c + \Delta T_t + \Delta T_r + \Delta T_k = \Delta T_c + Iv(\frac{rR}{2h})\frac{\Delta H}{C_p} + \frac{2\Gamma_{SL}}{r} + \frac{R}{\mu} \quad (7)$$

where $h$ is thermal diffusivity, $\Delta H$ is the latent enthalpy of SS316L, $C_p$ is the specific heat capacity, $\Gamma_{SL}$ is the Gibbs−Thomson coefficient, and $\mu$ is the interface kinetic coefficient. The radius of dendrite tip $r$ is obtained by solving the following equation,

$$4\pi^2\Gamma_{SL}(\frac{1}{r^2})+\left\{2\sum_i\left[m_{v,i}P_{e,i}(1-k_{v,i})C_i^*\xi_{c,i}\right]\right\}(\frac{1}{r})+G=0 \tag{8}$$

$$\xi_{c,i}=1-\frac{2k_{v,i}}{2k_{v,i}-1+\left[1+(2\pi/p_{e,i})^2\right]^{1/2}} \tag{9}$$

$$P_{e,i}=\frac{rR}{2D_i} \tag{10}$$

where $m_{v,i}$ is the dynamic liquid slope, $P_{e,i}$ is the solute Peclet, $k_{v,i}$ is the dynamic partition coefficient related to solidification rate, and $C_i^*$ is the composition concentration at the dendrite tip. The subscript $i$ represents the element.

The constitutional undercooling $\Delta T_c$ can be expressed as [7],

$$\Delta T_c=\sum_{i=1}^{q}(m_{v,i}C_i^*-m_{0,i}C_{0,i})=\sum_{i=1}^{q}C_{0,i}\left[\frac{m_{v,i}}{1-(1-k_{v,i})Iv(p_{e,i})}-m_{0,i}\right] \tag{11}$$

$$k_{v,i}=\frac{k_{0,i}+\delta R/D_i}{1+\delta R/D_i} \tag{12}$$

$$m_{v,i}=m_{0,i}\left[\frac{1-k_{v,i}\left(1-\log(k_{v,i}/k_{0,i})\right)}{1-k_{0,i}}\right] \tag{13}$$

where $C_{0,i}$ is the nominal composition of component $i$, $k_{0,i}$ is the equilibrium partition coefficient for element $i$, $D_i$ denotes the liquid-phase diffusivity calculated from the DICTRA module of Thermo-Calc, where multicomponent diffusion interactions among all alloying elements are accounted for at the local-liquid interface temperature. $m_{0,i}$ is the equilibrium liquidus slope of SS316L element via pseudo-binary phase diagram. The interface width $\delta$ is assumed to be 1.0 nm [20]. The solute concentration distribution in the mushy zone is approximately as follows [32]

$$C_i^L=C_{0,i}+(C_{0,i}/k_{v,i}-C_{0,i})\exp(-x/l_i^d) \tag{14}$$

where $x$ is the distance from the solidus phase line along the direction of solidification, and $l_i^d=D_i/R$ is the diffusion length, representing the characteristic distance over which solute redistribution occurs.

Table 1 Material parameters used in the simulation of SS316L

| Parameters | Symbol | Unit | Value |
|---|---|---|---|
| Solid-liquid interfacial energy | $\gamma$ | $J/m^2$ | 0.5 |
| Density of potential nucleation site | $N_0$ | $1/m^3$ | $2.0\times10^{15}$ |
| Equiaxed exponent | $n$ | 1 | 3.4 |
| Thermal diffusivity | $h$ | $m^2/s$ | $3.6\times10^{6}$ |
| Nucleation undercooling | $\Delta T_n$ | K | 4.0 |
| Fusion entropy of per volume | $\Delta S_f$ | J/m$^3$/K | 1.04×10$^6$ |
| Latent enthalpy | $\Delta H$ | kJ/kg | 260 |
| Gibbs−Thomson coefficient | $\Gamma$ | K·m | 2.4×10$^{-7}$ |
| Size of the nucleating site | $d$ | μm | 0.5 |
| Interface kinetic coefficient | $\mu$ | m/s/k | 0.4 |
| Thermodynamic database | / | / | TCFE13 |
| Mobility database | / | / | MOBFE8 |
| Diffusion coefficient of Cr | $D_{L,Cr}$ | m$^2$/s | 3.0×10$^{-9}$ |
| Diffusion coefficient of Mo | $D_{L,Mo}$ | m$^2$/s | 3.5×10$^{-9}$ |
| Diffusion coefficient of Si | $D_{L,Si}$ | m$^2$/s | 3.6×10$^{-9}$ |
| Diffusion coefficient of Ni | $D_{L,Ni}$ | m$^2$/s | 3.3×10$^{-9}$ |
| Diffusion coefficient of S | $D_{L,S}$ | m$^2$/s | 2.3×10$^{-9}$ |
| Diffusion coefficient of P | $D_{L,P}$ | m$^2$/s | 2.1×10$^{-9}$ |
| Diffusion coefficient of C | $D_{L,C}$ | m$^2$/s | 4.7×10$^{-9}$ |

Table 2. SS316L composition

| Element | Fe | Cr | Ni | Mo | Si | C | P | S |
|---|---|---|---|---|---|---|---|---|
| Wt. % | Bal. | 17.0 | 12.0 | 2.5 | 1.0 | 0.03 | 0.04 | 0.03 |

Table 3. Liquid slope of SS316L element via pseudo-binary phase diagram

| Element | Cr | Ni | Mo | Si | C | P | S |
|---|---|---|---|---|---|---|---|
| $m_{0,i}$ (K/wt. %) | -3.88 | -1.025 | -5.18 | -10.16 | -66.6 | -34 | -35.5 |

2.3 Equiaxed fraction prediction

To establish a continuous thermodynamic description of CET behavior over broad solidification conditions, the CALPHAD-generated equiaxed fraction $\phi$ as a function of $G$ and $R$ was interpolated using GPR, as schematically illustrated Fig. 2a. For each alloy composition, discrete CALPHAD-predicted CET data ($G$, $R \rightarrow \phi$) were used to describe the relationship between solidification conditions and equiaxed grain fraction (Fig. 2b). A Matern kernel with $\nu = 0.5$ was employed to capture the inherently abrupt CET transition behavior under rapid solidification conditions. In this case, the kernel reduces to the exponential form,

$$k(x_i, x_j) = \exp(-\,d(x_i, x_j)/l \tag{15}$$

where $d(x_i, x_j)$ is the Euclidean distance between two input points in the ($G$, $R$) space, and $l$ is the characteristic length scale controlling the correlation between data points. The resulting interpolation provides a continuous CET description across broad solidification conditions within the melt pool (Fig. 2c), enabling thermodynamically consistent incorporation of local nucleation behavior into PF simulations. The consistency of the interpolated CET behavior is further demonstrated in Fig. 2(d), where close agreement between GPR-interpolated and CALPHAD-calculated equiaxed fractions is observed. The predicted $\phi$ value ranges from 0 and 1 under dendritic growth conditions, while negative values are used to represent planar growth regimes where equiaxed nucleation is suppressed.

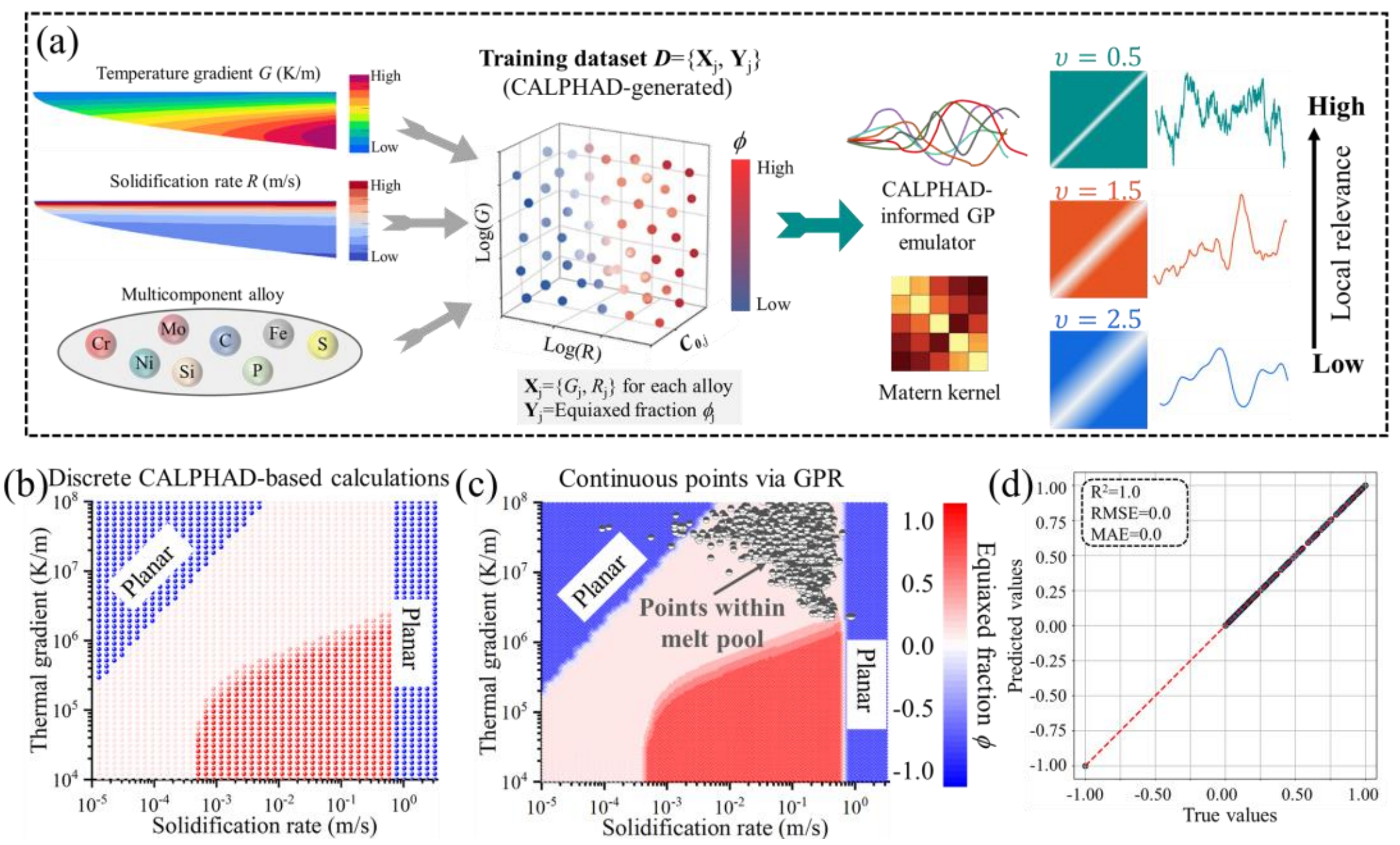

Fig. 2 (a) GPR-assisted continuous CET description relating temperature gradient $G$, solidification rate $R$, alloy composition and equiaxed fraction $\phi$. (b) Discrete CALPHAD-predicted equiaxed fractions under selected $G$-$R$ conditions. (c) Continuous CET map reconstructed through GPR. (d) Comparison between GPR-interpolated and CALPHAD-calculated equiaxed fractions.

## 2.4 CALPHAD-informed nucleation model

The prediction of nucleation probability within the melt pool is a critical factor for accurately capturing CET during rapid solidification. Conventionally, nucleation is described using statistical distribution functions characterized by a mean undercooling and standard deviation. However, these parameters are typically determined under near-equilibrium conditions and are highly sensitive to rapid thermal fluctuations [16], limiting their applicability to the highly non-equilibrium conditions encountered in LPBF.

To overcome these limitations, we adopt an alternative strategy by establishing a physically consistent link between the equiaxed fraction $\phi$ and nucleation probability $p$. Instead of directly prescribing nucleation probability, the local equiaxed fraction $\phi$ is obtained from CALPHAD-based CET calculations using Thermo-Calc, which inherently accounts for the coupled effects of thermal conditions and alloy composition under non-equilibrium solidification conditions. The equiaxed fraction is described using an Avrami-type formulation, in which nucleation dominates at early stages, while subsequent grain growth and impingement are implicitly accounted for. According, the equiaxed fraction $\phi$ can be expressed as [33],

$$\phi = 1 - \exp(-N_{act} A_g) \tag{16}$$

$$N_{act} = N_0^{2D} P_{cum}(t) \tag{17}$$

where $N_{act}$ denotes the cumulative activated nucleation density, $A_g$ represents effective area contribution of a single equiaxed grain, which is approximated by the initial nucleus area adopted in PFM, $N_0^{2D}$ is the equivalent two-dimensional density of potential nucleation sites for the present two-dimensional PF simulation, and $P_{cum}$ is the cumulative activation probability at time $t$.

The nucleation probability $p$ at each lattice site within a given update interval $\Delta t$ can be expressed as,

$$p = P_{cum}(t) - P_{cum}(t - \Delta t) \tag{18}$$

Substituting Eqs. (16) and (17) into Eq. (18) yields

$$p = -\frac{1}{N_0^{2D} A_g} \ln\left[\frac{1-\phi(t)}{1-\phi(t-\Delta t)}\right] \tag{19}$$

This formulation ensures that nucleation is introduced incrementally during the simulation and that each potential nucleation site is activated at most once, providing a consistent bridge between CET-predicted equiaxed fraction and stochastic nucleation in PF simulation.

## 3. Results and discussion

### 3.1 Limitations of conventional nucleation model under rapid solidification

Fig. 3 compares the spatial distribution of nucleation probability along the melt pool boundary predicted by the conventional nucleation model and the CALPHAD-informed approach under the power of 195 W and scanning speed of 1.083 m/s. The corresponding thermal conditions, characterized by the spatial distributions of thermal gradient $G$ and solidification rate $R$, are obtained from thermal simulations following our previous work [27, 30]. The conventional nucleation model produces a smooth distribution with a peak near the top region of the melt pool, where nucleation behavior is prescribed through empirically defined Gaussian parameters, including the mean nucleation undercooling $\mu$ and standard deviation $\sigma$.

In contrast, the CALPHAD-informed nucleation model predicts nearly zero nucleation at the top region, followed by a sharp increase and subsequent decrease along the melt pool boundary. The high solidification rate near the top region of melt pool enhances solute trapping and suppresses constitutional instability, thereby stabilizing planar growth and suppressing equiaxed nucleation. As the local thermal conditions evolve along the melt-pool boundary, the competition between constitutional undercooling and nonequilibrium partitioning leads to strong spatial variations in nucleation behavior. Such nonequilibrium solidification behavior cannot be captured by conventional Gaussian-based nucleation models due to their reliance on predefined empirical undercooling distributions that are not directly coupled with local multicomponent thermodynamic conditions. By incorporating CALPHAD-derived thermodynamic information, the present approach naturally reproduces the transition in nucleation behavior and captures the strong spatial variations inherent to LPBF solidification conditions.

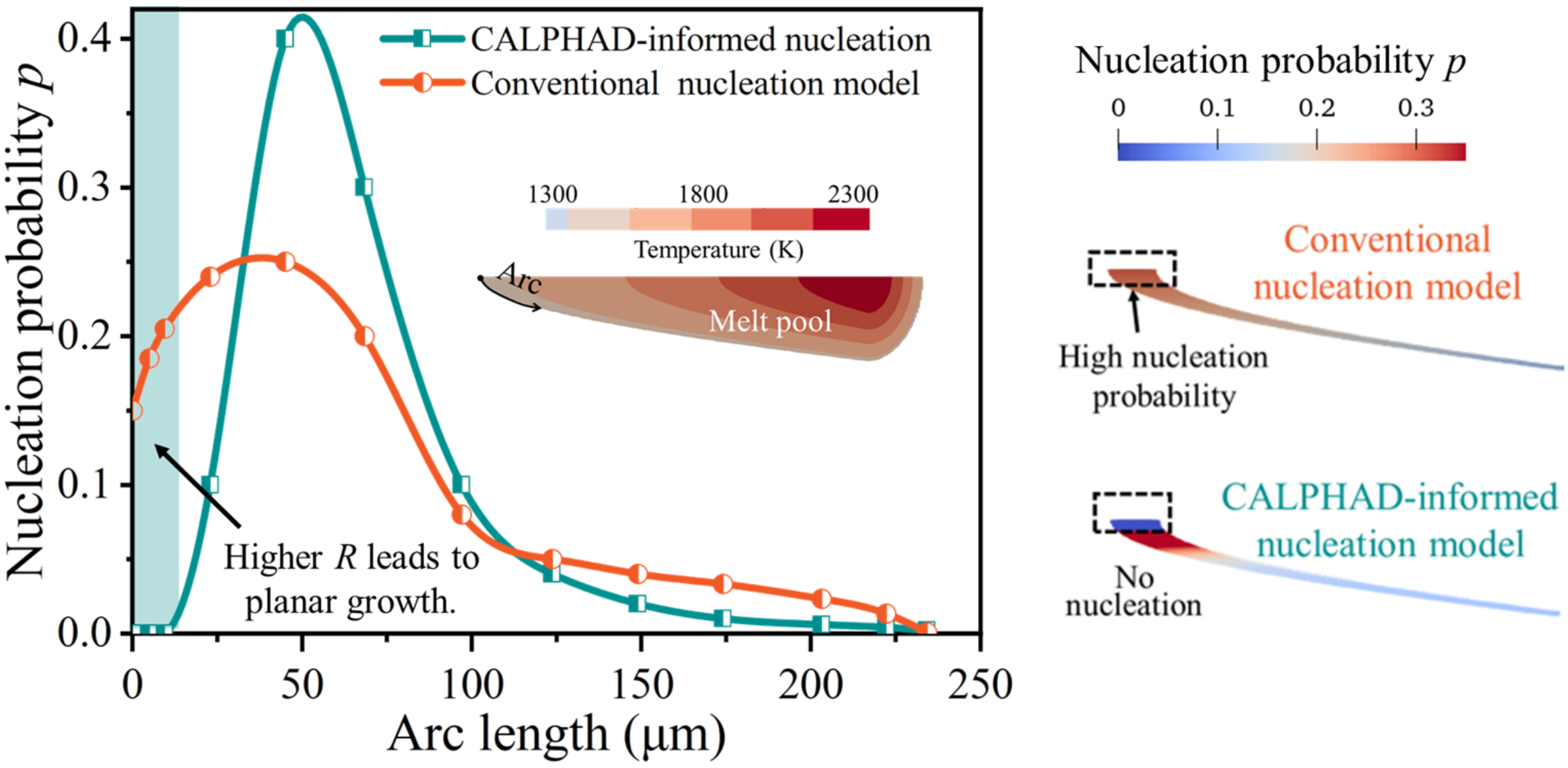

Fig. 3 Comparison between CALPHAD-informed and conventional nucleation models during LPBF solidification. The CALPHAD-informed nucleation model captures the strong local variations in $G$ and $R$ inherently associated with LPBF melt-pool dynamics, leading to spatially varying nucleation behavior under nonequilibrium solidification conditions.

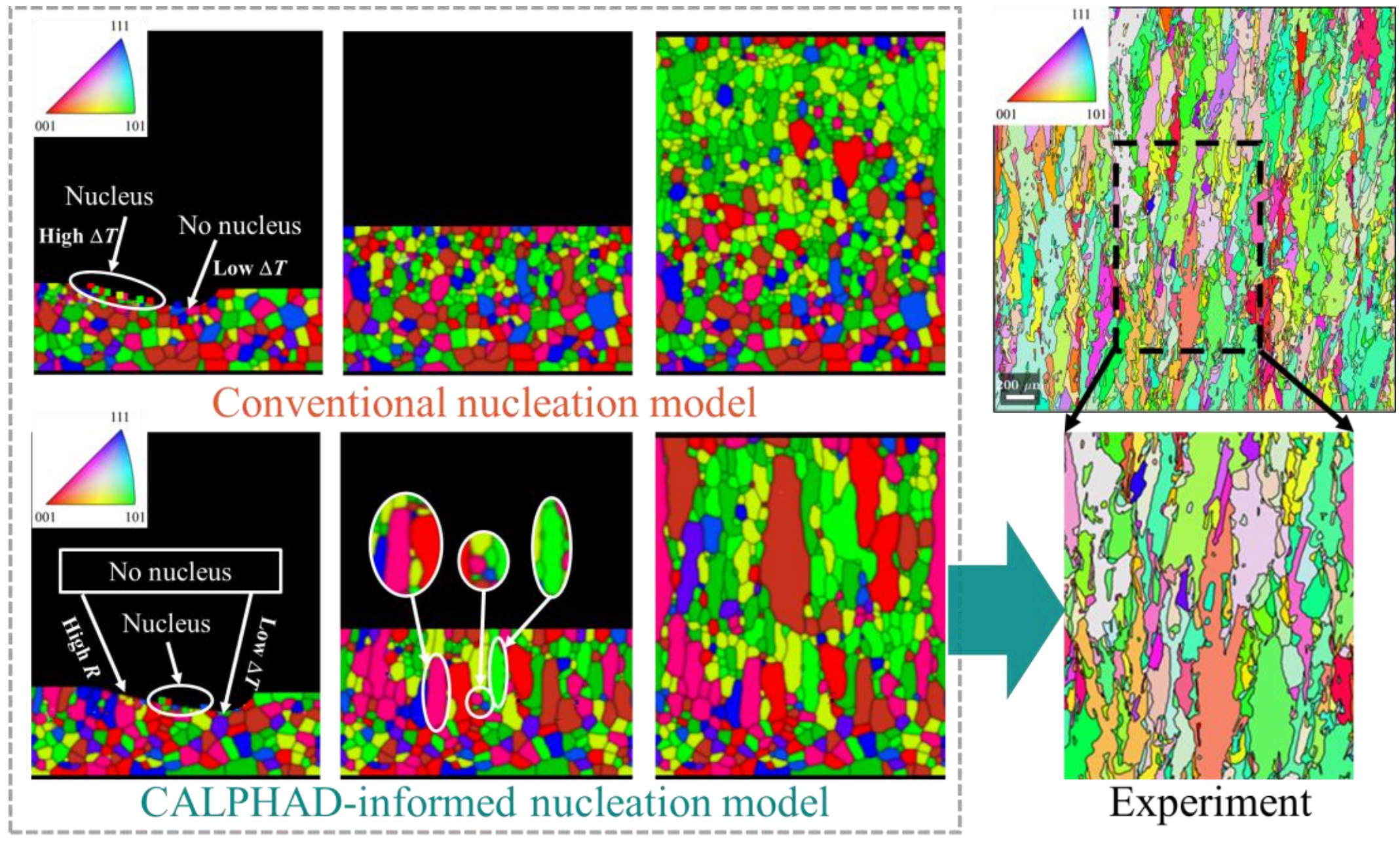

Fig. 4 Grain structure comparison between PF simulations using conventional and CALPHAD-informed nucleation models

Fig. 4 compares the grain structures predicted by PF simulation using the conventional and CALPHAD-informed nucleation models. The high nucleation probability near the melt pool top predicted by the conventional Gaussian-based model promotes widespread equiaxed grain formation, thereby suppressing the development of columnar structures. By contrast, the CALPHAD-informed nucleation model captures the strong spatial variation of nucleation behavior governed by local solidification conditions. Nucleation is suppressed near the top of melt pool due to enhanced solute trapping at high solidification rates as demonstrated in Fig. 3, and near the bottom region due to insufficient constitutional undercooling. Consequently, nucleation is confined to intermediate regions, allowing columnar grains to grow preferentially and dominate the microstructure. EBSD characterization further confirms the columnar-dominated grain structure predicted by the PF simulation using CALPHAD-informed nucleation model.

### 3.2 Origin of solute trapping in multicomponent alloys

Such variations in nucleation behavior are closely associated with non-equilibrium effects induced by high solidification rates, particularly solute trapping at the solid–liquid interface. Fig. 5 illustrates the mechanism of solute trapping, with carbon selected as a representative element due to its large liquids slope and strong segregation tendency at low concentration, as shown in Fig. 5(a). The solute concentration profile ahead of the solid-liquid interface follows an exponential decay, as shown in Fig. 5(b). As $R$ increases, the concentration gradient becomes steeper and the effective diffusion length decreases, indicating that long-range solute diffusion becomes increasingly suppressed. Consequently, solute atoms cannot be efficiently redistributed into the liquid ahead of the interface. At the same time, the degree of solute enrichment ahead of the interface decreases due to velocity-dependent partition coefficient changes, as shown in Fig. 5(c). At low $R$, solute diffusion dominates the redistribution process. However, as $R$ exceeds ~0.02 m/s, the partition coefficient $k_{v,C}$ increases sharply, indicating a transition from solute diffusion-controlled to interface kinetics-controlled solidification.

Fig. 5(d) compares the solute concentration at the dendrite tip $C_{tip}$ calculated with and without solute trapping. As $R$ exceeds ~0.02 m/s, the trapped carbon increases significantly, confirming the enhanced solute trapping under rapid solidification conditions. For example, the trapped concentration, defined as the difference in $C_{\mathrm{tip}}$ with and without solute trapping, increases from 0.0004 wt% at $R$=0.03 m/s to 0.077 wt% at $R$=0.8 m/s. This concentration variation further reduces the constitutional undercooling. As shown in Fig. 5(e), at $R$=0.8 m/s, the constitutional undercooling associated with carbon decreases from 13.62 K to 8.89 K as a result of solute trapping, corresponding to a reduction of 34.7%, highlighting the strong impact of solute trapping on suppressing constitutional undercooling. Considering the combined effect of all alloying elements, Fig. 5f compares the total undercooling and constitutional undercooling with and without solute trapping. When $R$ < 0.02 m/s, the constitutional undercooling $\Delta T_{\mathrm{c}}$ remains nearly unchanged. As $R$ increases to 0.8 m/s, $\Delta T_{\mathrm{c}}$ decreases from 50.4 K to 29.3 K, corresponding to a reduction of 41.8%, demonstrating the collective impact of multi-element solute trapping on the overall constitutional undercooling behavior.

In addition, solute redistribution affects the dendrite tip radius and the associated solute Peclet number, thereby influencing the curvature undercooling $\Delta T_r$ as described in Eqs. (7-10). As shown in Fig. 5f, the total undercooling $\Delta T_{total}$ reaches a maximum value of 33.6 K at $R$=0.35 m/s and then decreases with further increase in $R$ due to solute trapping. Solute trapping weakens the solute diffusion boundary layer ahead of the interface (Fig. 5b) and reduces solute partitioning (Fig. 5c), thereby suppressing solute-driven constitutional instability. As a result, the dendrite tip becomes less constrained by solute diffusion and tends to adopt a larger radius at high $R$, which reduces $\Delta T_r$ (Eq. 7) and leads to a decrease in $\Delta T_{total}$.

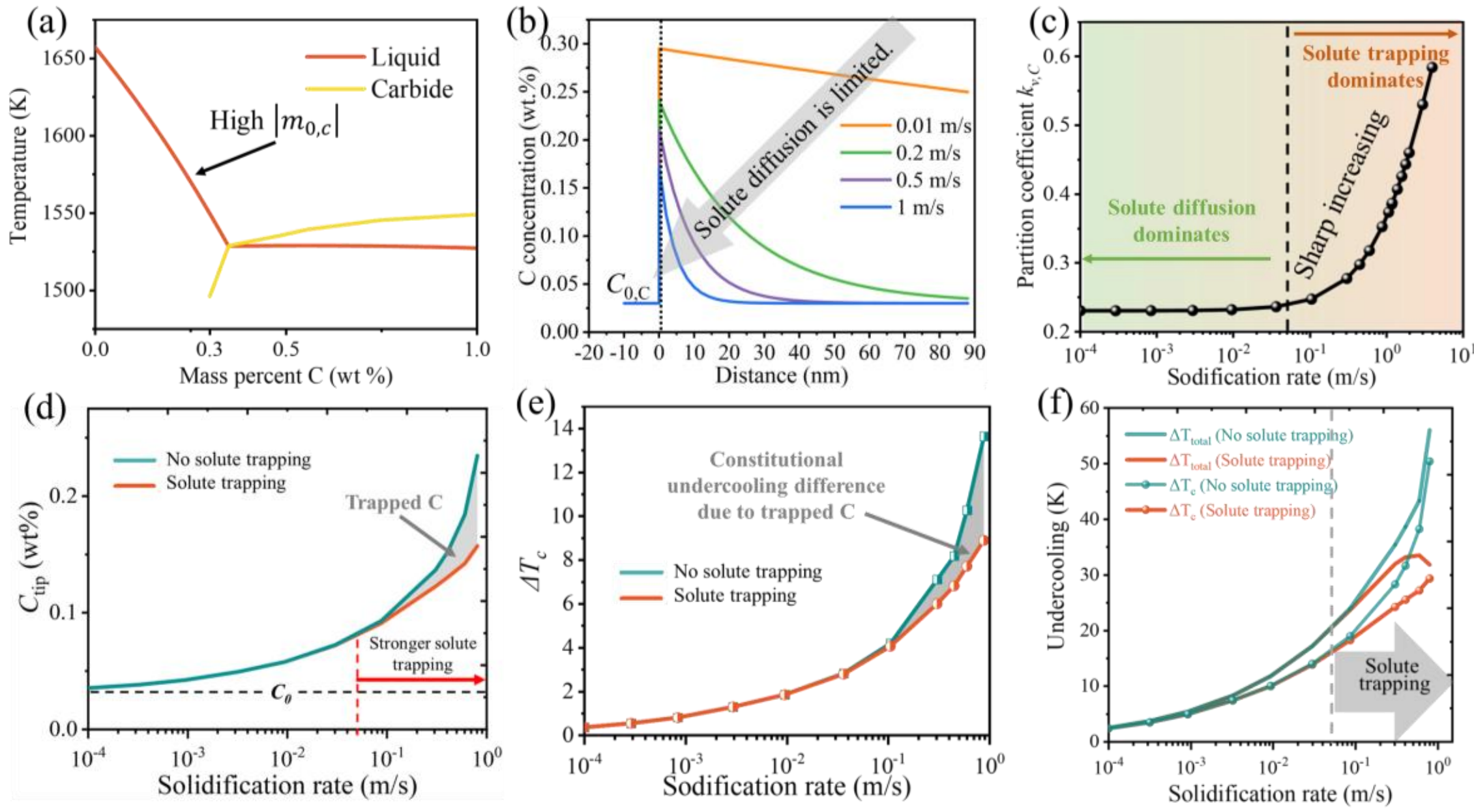

Fig. 5 Origin of solute trapping illustrated by carbon: transition from solute-diffusion to solute-trapping growth. (a) Equilibrium phase diagram showing liquid slope; (b) Evolution of solute concentration along the solidification direction; (c) Velocity-dependent partition coefficient; (d) Solute concentration at the dendrite tip; (e) carbon-induced constitutional undercooling; (f) Total undercooling and constitutional undercooling.

We further investigate the element-specific contributions to constitutional undercooling, as shown in Fig. 6. Different alloy elements exhibit distinct diffusivities, equilibrium partition coefficients, and liquidus slopes (as listed in Tab. 1-3), which together determine their sensitivity to nonequilibrium partitioning under rapid solidification conditions. Fig. 6(a) compares the contribution of individual alloying element to the constitutional undercooling over a wide range of solidification rates. At low solidification rates ($R$<0.24 m/s), the contribution of carbon is relatively small compared to that of Cr and Mo, primarily due to its low concentration. In this regime, Cr and Mo dominate the constitutional undercooling owing to their higher concentrations and significant partitioning behavior. As $R$ increases, the contribution of C rises and exceeds that of Mo at $R$>0.24 m/s and that of Cr at $R$>0.68 m/s. Specifically, the carbon contribution increases from 1.32 K at $R$=0.01 m/s to 3.02 K at $R$=0.09 m/s and 8.9 K at $R$=0.8 m/s. Correspondingly, its relative contribution increases from 13.2% to 16.6% and 30.4%, indicating that carbon becomes increasingly important to the overall constitutional undercooling at higher solidification rates.

To further elucidate the sensitivity of different alloying elements to solute trapping, the contribution of each element is decomposed into the actual value under non-equilibrium condition (with solute trapping) and the deviation induced by solute trapping from equilibrium at a solidification rate of $R$=0.8 m/s, as shown in Fig. 6 (b). The normalized deviation relative to the equilibrium value provides a direct quantitative measure of the element-specific sensitivity to solute trapping. The deviation for C, Cr, and Mo is 4.7, 2.17, and 1.98 K, corresponding to 34.6%, 21.8%, and 28.9% of their respective equilibrium values (13.6, 9.97, and 6.86 K), indicating a relatively moderate sensitivity to solute trapping. This behavior can be attributed to their comparatively higher equilibrium partition coefficients and diffusivities. Despite this moderate sensitivity, C, Cr, and Mo provide the largest contributions to constitutional undercooling, with values of 8.9, 7.8, and 4.88 K, respectively, corresponding to approximately 30.4%, 26.6%, and 16.7% of the total constitutional undercooling (29.3 K), indicating their dominant role in determining the local solidification response.

In contrast, the deviation for S and P are 7.4 and 4.75 K, corresponding to 66.7% and 57.9 % of their respective equilibrium values (11.2 and 8.2 K), showing a much stronger sensitivity to solute trapping. Owing to their extremely low equilibrium partition coefficients ($k_0$<0.1) and limited atomic diffusivities, the effective partitioning of these elements is highly sensitive to interface kinetics. As a result, the solute trapping effect is promoted, which suppresses solute buildup ahead of the solid–liquid interface and substantially reduces their contribution to constitutional undercooling.

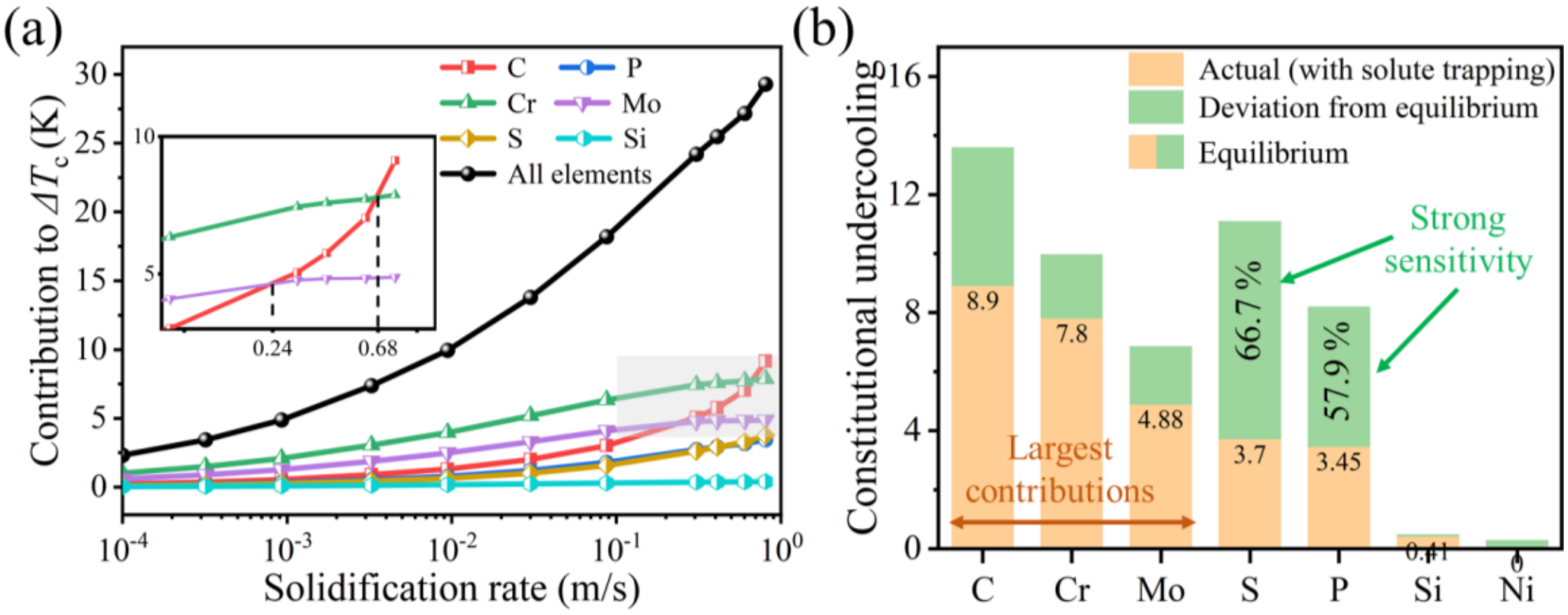

Fig. 6 Effect of alloying elements on constitutional undercooling. (a) Element-resolved contributions to constitutional undercooling; (b) Decomposition of the constitutional undercooling into the actual contribution under non-equilibrium conditions and the deviation induced by solute trapping for individual alloying elements at $R$=0.8 m/s. The normalized deviation relative to the equilibrium value provides a quantitative measure of the element-specific sensitivity to solute trapping.

3.2.1 Atom diffusivity

To identify the factors governing solute trapping, a parametric study is conducted by varying the liquid diffusivity of the alloying element, illustrated using carbon. Fig. 7(a) shows the solute concentration distribution along the solidification direction at a fixed $R$ of 0.5 m/s for representative diffusivities of $2.0\times10^{-9}$, $4.7\times10^{-9}$, and $1.0\times10^{-8}$ $m^2/s$. Increasing diffusivity results in a broader solute diffusion profile ahead of the solid–liquid interface, indicating enhanced solute redistribution in the liquid. The effect of diffusivity on nonequilibrium partitioning is further illustrated in Fig. 7(b). Lower diffusivity leads to a more rapid increase in the velocity-dependent partition coefficient $k_{v,C}$, indicating stronger solute trapping.

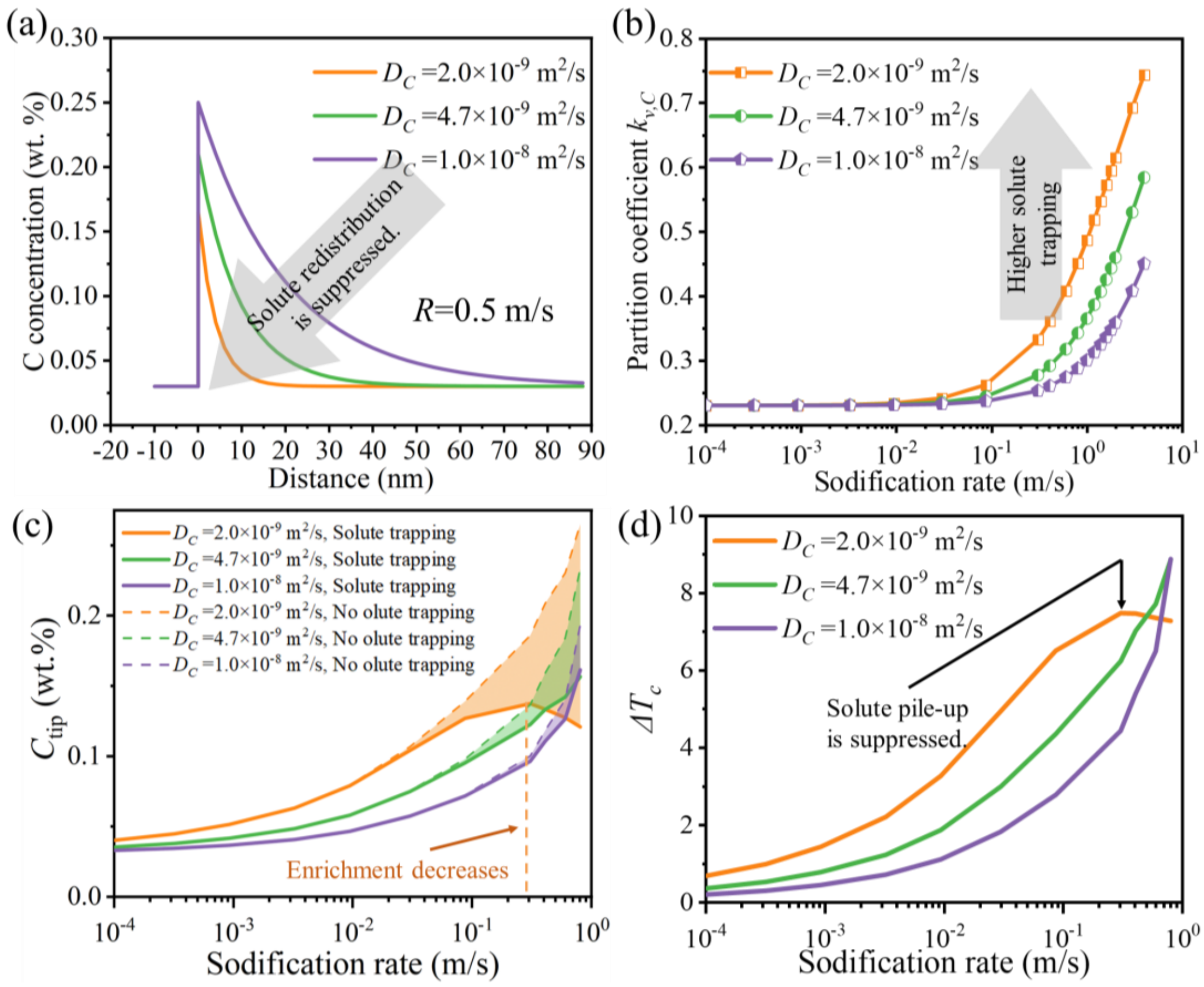

Fig. 7 Effect of diffusivity illustrated by carbon on (a) solute concentration distribution along the solidification direction; (b) velocity-dependent partition coefficient; (c) solute concentration at the dendrite tip front; (d) carbon-induced constitutional undercooling.

Diffusivity is directly related to the solute Peclet number (Eq. 10) and thus affects the solute concentration in the liquid at the dendrite tip front $C_{tip}$. As shown in Fig. 7 (c), at low diffusivity ($2.0\times10^{-9}$ $m^2/s$), limited solute mobility suppresses diffusion away from the interface, leading to localized solute accumulation and an elevated $C_{tip}$. at low solidification rate $R$. The increase in $C_{tip}$ lowers the local equilibrium liquidus temperature and thereby enhances the constitutional undercooling, as shown in Fig. 7(d). However, for low diffusivity, as $R$ increases further (>0.16 m/s), enhanced solute trapping suppresses solute enrichment ahead of the interface (as indicated in Fig. 7c), leading to a reduction in $C_{tip}$ and, consequently, a decrease in constitutional undercooling. This reflects a transition from a solute diffusion-controlled regime at low solidification rates to a solute trapping-dominated regime at high solidification rates.

### 3.2.2 Interface width δ

In addition to diffusivity, the interface width $\delta$, representing the characteristic length scale over which the solid–liquid transition occurs, also affects solute trapping. As shown in Fig. 8(a), higher $\delta$ leads to a higher velocity-dependent partition coefficient, indicating stronger solute trapping. As a result, the solute pile-up at the dendrite tip is reduced, as shown in Fig. 8(b).Fig. 8(c) further illustrates the effect of $\delta$ on carbon-induced constitutional undercooling under different solidification rates $R$. The influence of $\delta$ becomes more pronounced at higher $R$. For example, at $R$=0.8 m/s, decreasing $\delta$ decreases from 2.0 nm to 0.2 nm increases the constitutional undercooling from 7.11 K to 12.11 K, approaching the case without solute trapping.

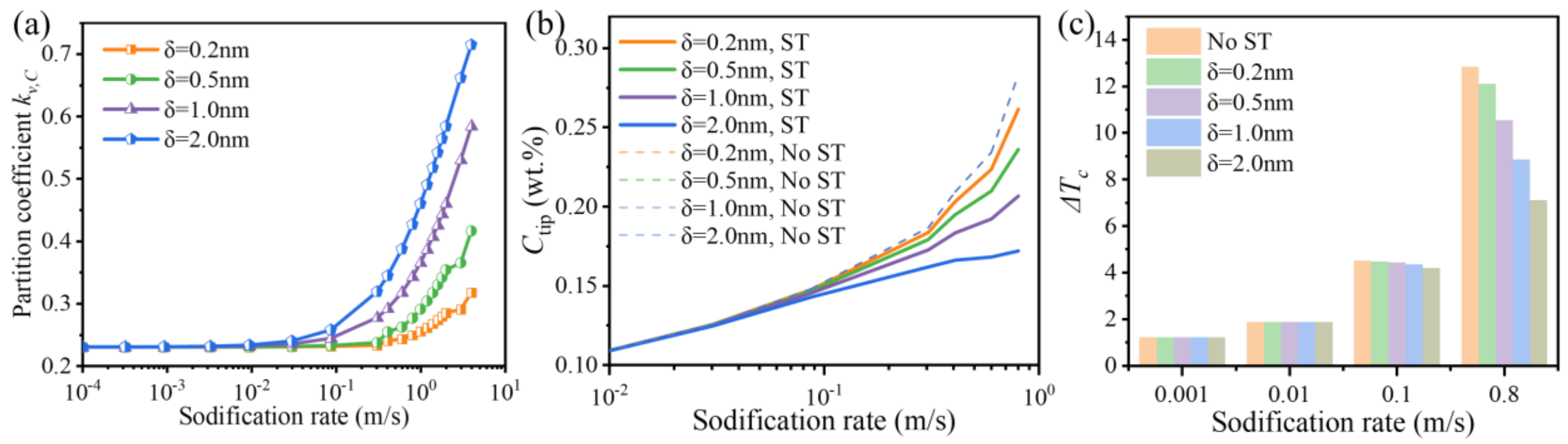

Fig. 8 Effect of interface thickness on (a) partition coefficient $k_{v,C}$, (b) dendrite tip concentration, and (c) contribution of carbon to constitutional undercooling.

## 3.3 Microstructure evolution governed by solute trapping and experimental validation

We perform a quantitative comparison of grain structure and grain size distribution between experimental EBSD observations and CALPHAD-informed PF simulations under three different volumetric energy densities (VED) $P/V/h/t$, where $h$ and $t$ represent the hatch spacing and layer thickness, respectively, as shown in **Error! Reference source not found.**. The simulated grain structures exhibit good agreement with the EBSD observation under selected three cases. Fig. 10 plots the quantitative verification of grain size and aspect ratio under different processing parameters, both of which follow a log-normal distribution. When VED increases from $P_1$ to $P_3$, the experimentally measured median equivalent grain size is 21.82, 24.65, and 33.96 μm, respectively, while the corresponding simulation results are 23.01, 24.17, and 35.10 μm, showing good agreement and the trend that high energy input leads to larger grain size, as shown in Fig. 10(a). Grain anisotropy is another key metric for characterizing grain morphology. To quantify this, the length-to-width ratio is evaluated. As shown in Fig. 10(b), the mean length-width ratios are 3.15, 2.91, and 3.5 for the experiment, and 3.12, 2.78, and 3.18 for the simulation. Overall, the CALPHAD-informed PF model demonstrates strong predictive capability in capturing both the grain size and anisotropy under LPBF conditions.

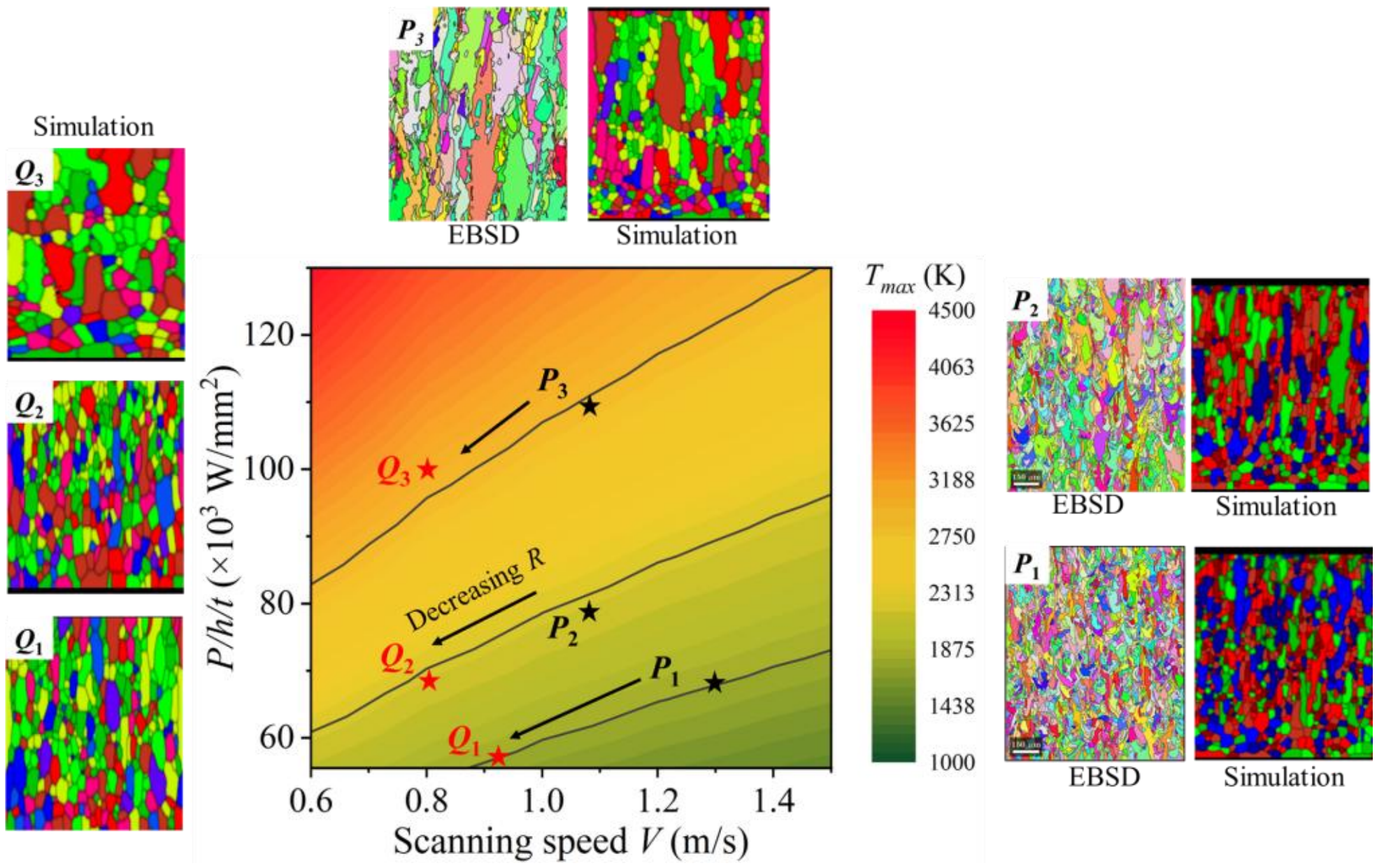

Fig. 9 Microstructure evolution between experiment and CALPHAD-informed PFM simulation under multiple processing conditions.

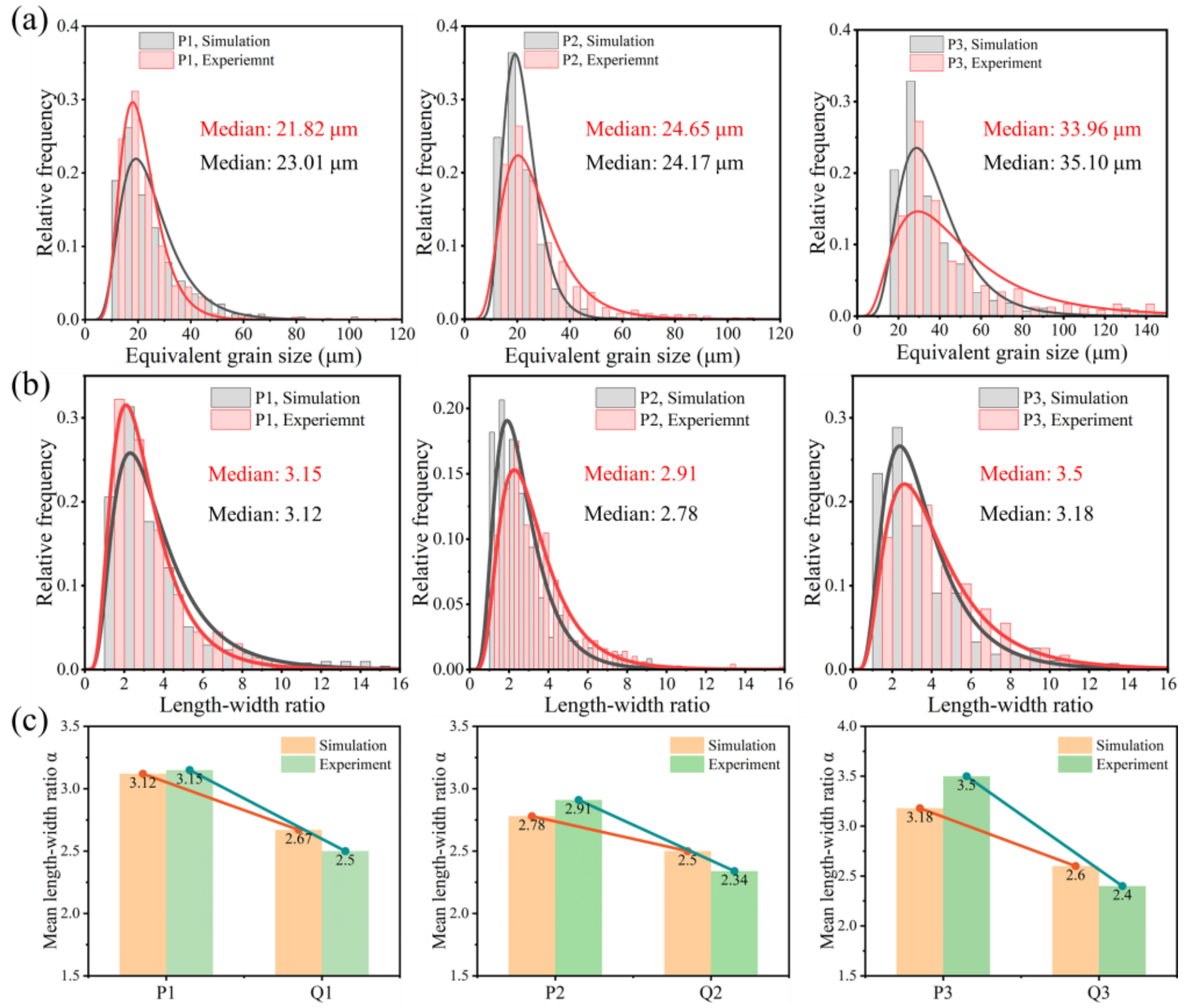

Fig. 10 Quantitative verification of grain size and aspect ratio under different processing parameters: (a) equivalent grain size; (b) length-width ratio; (c) length-width ratio comparison under different scanning speeds $V$ at comparable VED.

To validate the effect of solidification rate-dependent solute trapping effect on microstructure evolution, the length-width ratio under different scanning speeds $V$ at comparable VED is compared, as shown in Fig. 10(c). As $V$ decreases from $P_1$ to $Q_1$, the simulated length-width ratio decreases from 3.12 to 2.67, in good agreement with experimental measurements, which decrease from 3.15 to 2.5 [34]. Similarly, for $P_2$ to $Q_2$, the simulated ratio decreases from 2.78 to 2.50, compared to 2.91 to 2.34 in experiments [35]. For $P_3$ to $Q_3$, the simulated values decrease from 3.18 to 2.60, while the experimental values decrease from 3.50 to 2.40. Overall, a decrease in solidification rate leads to a reduction in the length-to-width ratio, indicating a transition from elongated columnar grains toward more equiaxed morphologies. This behavior can be attributed to the reduced extent of solute trapping at lower solidification rates, which promotes stronger solute partitioning and weakens the directional growth stability.

Fig. 11 plots the transition of grain morphology, characterized by the length-width ratio α, undercooling, and nucleation probability as a function of solidification rate $R$. A temperature gradient of $2.0\times10^6$ K/m is adopted as a representative value within the typical LPBF range for evaluating the nucleation probability. As $R$ increases, the total undercooling exhibits a transition from being dominated by constitutional undercooling $\Delta T_c$ to kinetic undercooling $\Delta T_k$. At low $R$ (<0.09 m/s), solute diffusion is sufficient to maintain near-equilibrium partitioning, resulting in pronounced $\Delta T_c$. However, due to the limited total undercooling, the driving force for nucleation remains insufficient. As a result, nucleation is suppressed, and grain growth is dominated by stable directional solidification, leading to the formation of coarse columnar structures (α>4).

As $R$ increases (0.09 m/s<$R$<0.4 m/s), the total undercooling becomes sufficient to activate nucleation. The enhanced driving force promotes grain formation ahead of the solid–liquid interface, resulting in a transition toward finer microstructures with a mixed equiaxed–columnar morphology (2.2<α<3.5). With $R$ increasing further (0.4 m/s< $R$<0.8 m/s), the solute trapping becomes pronounced. As a result, the growth of $\Delta T_c$ with respect to $R$ becomes increasingly limited. At the same time, the solidification mechanism shifts from solute diffusion-controlled to solute trapping-controlled growth, which weakens the concentration gradient and suppresses interfacial instability, leading to a blunted dendrite tip accompanied by a gradual decrease in radial undercooling $\Delta T_r$.

When $R$ exceeds the absolute stability velocity $V_a$ (0.8 m/s), solute trapping becomes dominant, marking a transition from diffusion-controlled to interface-controlled solidification. In this regime, solute partitioning is effectively suppressed, resulting in a diminished contribution from constitutional undercooling $\Delta T_c$, while kinetic undercooling $\Delta T_k$ becomes the primary contributor to the total undercooling. As a result, microstructure evolution is dominated by interface stability, leading to elongated columnar or near-planar morphologies.

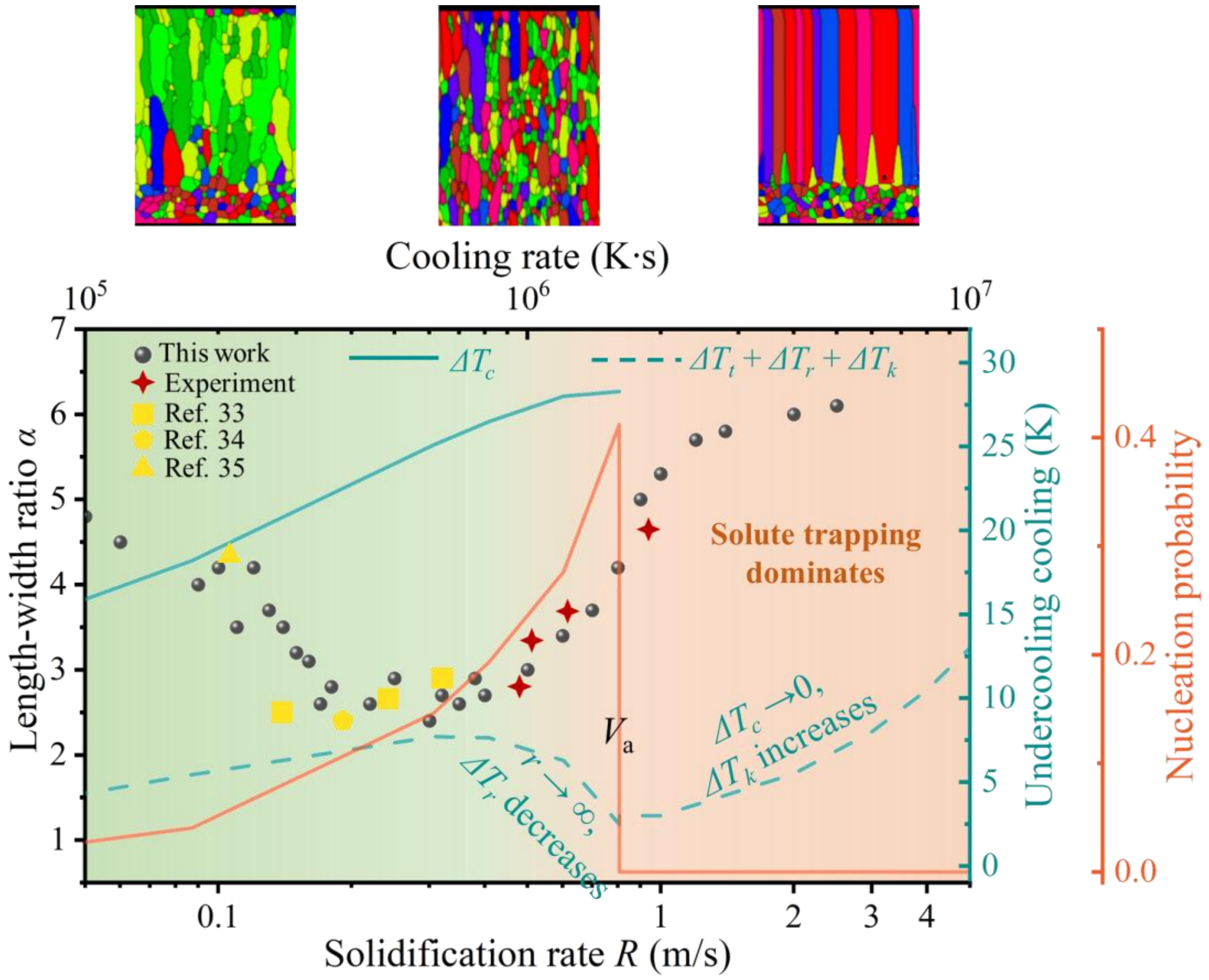

Fig. 11 Transition in grain morphology, undercooling, and nucleation probability as a function of solidification rate. A temperature gradient of $2.0\times10^6$ K/m is adopted as a representative value within the typical LPBF range for evaluating the nucleation probability.

### 3.4 Composition-dependent solute trapping and microstructure evolution

To gain further insight into composition-dependent effects, the influence of Cr content on solute trapping and microstructure evolution is examined under the laser power of 150 W and scanning speed of 1.1 m/s. The thermal conditions, characterized by the solidification rate $R$ and temperature gradient $G$, within the mushy zone are shown in Fig. 12(a), while representative ($G$, $R$) points are mapped onto CALPHAD-derived solidification microstructure selection maps under different Cr contents in Fig. 12(b). On the one hand, increasing Cr content leads to higher constitutional undercooling, which promotes nucleation over a broader region under low $R$ ($<V_a$). On the other hand, the compositional variation influences interfacial stability, resulting in a shift of $V_a$ toward higher values, as shown in Fig. 12(b). This indicates that compositional variations simultaneously affect the undercooling and the absolute stability velocity, thereby altering the spatial distribution of solidification regimes within the melt pool.

The corresponding nucleation probability distribution within the mushy zone under different Cr content, obtained using CALPHAD-informed nucleation model, are shown in Fig. 12(c). With increasing Cr content, the region with high nucleation probability progressively expands towards to the top of the melt pool, which can be attributed to the increase in $V_a$. In addition, higher undercooling increases the thermodynamic driving force for nucleation, resulting in the formation of more nuclei. The grain size evolution induced by variations in Cr content is quantified in Fig. 12(d), where the equivalent grain size decreases from 26.5 to 22.2 μm as the Cr content increases from 11 wt% to 23 wt%. This indicates that compositional variations promote grain refinement by enhancing undercooling and modifying interfacial stability under LPBF condition. It should be noted that the present analysis focuses on primary solidification behavior and does not explicitly consider the formation of secondary phases such as carbides. Although such phases may influence microstructure evolution, especially at later stages, the current framework captures the dominant mechanisms governing nucleation and grain structure selection during rapid solidification.

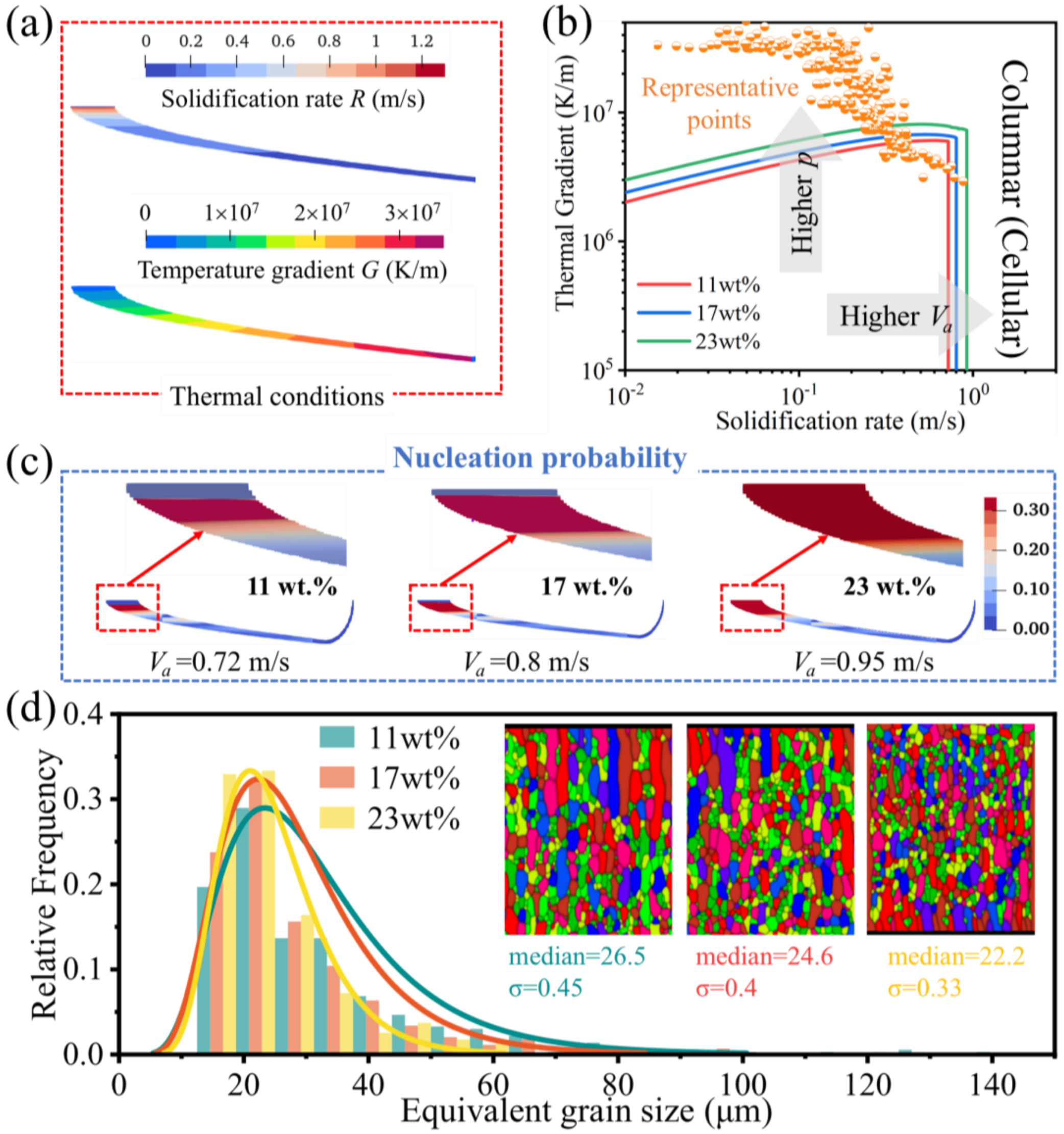

Fig. 12 Effect of Cr content on solute trapping and microstructure selection. (a) Thermal conditions within the mushy zone under a laser power of 150 W and a scanning speed of 1.1 m/s. (b) Microstructure selection map for different Cr contents, illustrating how Cr-induced changes in $V_a$, modify the stability boundary and thereby alter microstructure selection. (c) Nucleation probability distribution within the mushy zone. (d) Corresponding grain structure.

### 3.5 Limitations

The present study focuses on primary solidification and grain selection behavior during LPBF solidification of multicomponent alloys. The nucleation behavior is primarily described based on local thermodynamic conditions characterized by thermal gradient *G*, solidification rate *R*, and nonequilibrium partitioning behavior. Consequently, the effect of inclusions, oxide particles, or substrate surface conditions is not explicitly considered in the present model. In addition, secondary phase precipitation and subsequent solid-state transformations are beyond the scope of the present study due to the extremely short thermal cycle associated with rapid solidification during LPBF. Nevertheless, the interactions between second-phase evolution, grain size, and microstructural anisotropy may further influence grain selection behavior and therefore warrant future investigation.

Although powder bed geometry and melt-pool convection can influence local heat transport, grain nucleation and growth are primarily governed by the interfacial solidification conditions G and R, which are sufficiently captured by the present thermal model. Nevertheless, the present CALPHAD-informed phase-field simulations successfully reproduce the experimentally observed grain selection behavior under multiple LPBF processing conditions, indicating that the dominant nonequilibrium solidification physics governing nucleation behavior and grain structure evolution are reasonably captured.

## 4. Conclusion

In this work, CALPHAD-informed PF simulations were employed to investigate the effects of multicomponent thermodynamics and element-specific solute trapping on nucleation behavior and grain structure evolution during LPBF solidification of SS316L. By incorporating CALPHAD-based CET behavior into PF simulations, the coupled effects of thermal conditions (*G* and *R*), nonequilibrium partitioning, constitutional undercooling, and solute trapping on grain selection behavior were systematically analyzed and quantitatively validated against EBSD characterization under multiple LPBF processing conditions. The main conclusions can be summarized as follows:

1. The CALPHAD-informed nucleation model overcomes the limitations of conventional empirical nucleation models by incorporating multicomponent nonequilibrium thermodynamics into nucleation behavior prediction. Continuous CET description further enables thermodynamically consistent incorporation of composition-dependent nucleation behavior into PF simulations.
2. The contributions of individual alloying elements to constitutional undercooling evolve with solidification rate *R*, with C, Cr, and Mo dominating the overall undercooling at a high $R$=0.8 m/s, they account for 30.4%, 26.6%, and 16.7% of the total constitutional undercooling, respectively.
3. S and P exhibit strong sensitivity to solute trapping, with deviations from their equilibrium contributions to constitutional undercooling of 7.4 and 4.75 K at a solidification rate of $R$=0.8 m/s, corresponding to 66.7 % and 57.9 % of their equilibrium values, respectively; this pronounced sensitivity is attributed to their extremely low equilibrium partition coefficients and limited atomic diffusivities.

4. Nucleation behavior and resulting microstructures exhibit a strong dependence on solidification rate: when $R$<0.09 m/s, nucleation is suppressed due to insufficient undercooling, leading to coarse columnar structures; at intermediate $R$ (0.09<$R$<0.4 m/s), increased undercooling activates nucleation and promotes grain refinement; at higher $R$ (0.4<$R$<0.8 m/s), enhanced cooling rates promote solute trapping, which suppresses nucleation and results in higher grain anisotropy.
5. Compositional variations and solute trapping effects jointly govern microstructure evolution by modifying both constitutional undercooling $\Delta T_c$ and absolute stability velocity $V_a$, thereby controlling nucleation behavior within the mushy zone. Increased solute enrichment elevates $\Delta T_c$ and shifts $V_a$, enhancing nucleation probability and resulting in refined grain size.

Overall, the present study demonstrates the critical role of element-specific solute trapping and nonequilibrium thermodynamics in governing constitutional undercooling, nucleation behavior, and grain selection during LPBF solidification of multicomponent alloys. These findings provide fundamental insight into nonequilibrium grain selection and composition-dependent microstructure evolution during rapid solidification of multicomponent alloys.

### CRediT authorship contribution statement

**Xinxin Yao**: Conceptualization, Data curation, Formal analysis, Methodology, Software, Validation, Visualization, Writing – original draft, Writing – review & editing. **James Hanagan**: Data curation, Methodology, Software, Writing – review & editing. **Md Shafiqur Rahman Jame**: Data curation, Validation. **Mallikharjun Marrey**: Data curation, Validation. **Mohsen Taheri Andani**: Funding acquisition, Project administration. **Raymundo Arróyave**: Conceptualization, Investigation, Project administration, Resources. **Veera Sundararaghavan**: Investigation, Project administration, Resources, Supervision. **Lei Chen**: Funding acquisition, Investigation, Resources, Supervision, Project administration, Writing – review and editing.

### Declaration of Competing Interest

The authors declare that they have no known competing financial interests or personal relationships that could have appeared to influence the work reported in this paper.

### Declaration of generative AI use

During the preparation of this work, the authors used ChatGPT to improve the readability and clarity of the manuscript. After using this tool, the authors carefully reviewed and edited the content as needed and take full responsibility for the content of the publication.

**Acknowledgements**

This work was supported in part by the Defense Advanced Research Projects Agency (DARPA) SURGE program under Cooperative Agreement No. HR0011-25-2-0009, “Predictive Real-time Intelligence for Metallic Endurance (PRIME).